\magnification=\magstep1

\baselineskip 12pt
\parskip 4pt plus 1pt


\def\eg{{\it {\frenchspacing e.{\thinspace}g.  }}}

\def\gtsim{\mathrel{\raise.3ex\hbox{$>$}\mkern-14mu \lower0.6ex\hbox{$\sim$}}}

\def\ltsim{\mathrel{\raise.3ex\hbox{$<$}\mkern-14mu \lower0.6ex\hbox{$\sim$}}}


\font\lgb=cmbx10 scaled \magstep2

\font\lr=cmr10 scaled \magstep1


\null
\bigskip
\bigskip
\bigskip
\bigskip
{\lr Report of a Workshop on:}
\bigskip
\bigskip
\centerline{\lgb Smooth Particle Hydrodynamics:}
\bigskip
\centerline{\lgb Models, Applications, and Enabling Technologies}
\bigskip
\bigskip
\bigskip
\centerline{\lr Piet Hut$^1$ (chair)}
\bigskip
\centerline{\lr Lars Hernquist$^2$, George Lake$^3$, Jun Makino$^4$, Steve
McMillan$^5$, Thomas Sterling$^6$}
\smallskip
\centerline{\lr (scientific organizing committee)}
\bigskip
\bigskip
\bigskip
\bigskip
\bigskip
\bigskip
\bigskip
\bigskip
\noindent
The workshop was held on June 18-19, 1997, at the Institute for
Advanced Study.
\bigskip
\noindent
{\it Acknowledgments:}
This workshop was sponsored by NASA, through Dr. James Fischer,
Project Manager, NASA HPCC/ESS Project, and the Universities Space
Research Association, at Goddard Space Flight Center.
\bigskip
\bigskip
\item{$^1$}{Institute for Advanced Study, Princeton, NJ 08540 (piet@ias.edu)}
\smallskip
\noindent
\item{$^2$}{Lick Observatory, University of California,
Santa Cruz, CA 95064 $\qquad\qquad\qquad\qquad$
(lars@lick.ucolick.org)}
\smallskip
\noindent
\item{$^3$}{Astronomy Department, University of Washington,
Seattle, WA 98195 $\qquad\qquad\qquad$ (lake@astro.washington.edu)}
\smallskip
\noindent
\item{$^4$}{Department of General Systems Study,
College of Arts and Sciences, University of Tokyo, 3-8-1 Komaba,
Meguro-ku, Tokyo 153, Japan (makino@grape.c.u-tokyo.ac.jp)}
\smallskip
\noindent
\item{$^5$}{Department of Physics and Atmospheric Science,
Drexel University, Philadelphia, PA 19104 (steve@zonker.drexel.edu)}
\smallskip
\noindent
\item{$^6$}{High Performance Computing Systems Group, Jet Propulsion
Laboratory, Pasadena, CA 91109; Center for Advanced Computing
Research, California Institute of Technology, Pasadena, CA 91125
(tron@cacr.caltech.edu)}

\vfill\eject

\bigskip\noindent
{\bf Abstract}

\smallskip
\noindent
We present the results from a two-day study in which we discussed
various implementations of Smooth Particle Hydrodynamics (SPH), one of
the leading methods used across a variety of areas of large-scale
astrophysical simulations.  In particular, we evaluated the
suitability of designing special hardware extensions, to further boost
the performance of the high-end general purpose computers currently
used for those simulations.  We considered a range of hybrid
architectures, consisting of a mix of custom LSI and reconfigurable
logic, combining the extremely high throughput of Special-Purpose
Devices (SPDs) with the flexibility of reconfigurable structures,
based on Field Programmable Gate Arrays (FPGAs).

\smallskip
\noindent
The main findings of our workshop consist of a clarification of the
decomposition of the computational requirements, together with
specific estimates for cost/performance improvements that can be
obtained at each stage in this decomposition, by using enabling
hardware technology to accelerate the performance of general purpose
computers.

\smallskip
\noindent
The decomposition of astrophysical SPH applications is characterized
by four modules:

\item{1)}{
Most compute-intensive in any code is the module that computes the
pair-wise gravitational interactions;}

\item{2)}{
The naive $N^2$ force count, in a self-gravitating system of $N$
particles, can be reduced to $N\log N$ scaling, at the expense of
considerable computational overhead;}

\item{3)}{
The third-most computationally intensive module is the one performing
SPH;}

\item{4)}{
The remaining few percent of the total computational cost is taken
up by modeling additional physics, such as radiative transport and
other non-gravity/hydro effects.}

\noindent
For the specific case of large-scale cosmological simulations, a
leading application of SPH in astrophysics, we arrived at the
following estimates of cost/performance improvements that can be made
by utilizing a combination of SPD and FPGA accelerator technology.

\item{$\bullet$}{For $N^2$ gravity, a cost/performance improvement
of a factor 1000 can be achieved (and will be achieved in the
year 2000, by the 100 Tflops GRAPE-6 SPD, under development at the
University of Tokyo, through a grant from the Japanese Ministry of
Education).
}

\item{$\bullet$}{
More efficient $N\log N$ algorithms can be implemented as hardware
accelerators (\eg\ as SPD chips for the Ewald method, or an FPGA
subsystem for a Barnes-Hut tree), leading to a cost/performance
improvement of a factor 100.
}

\item{$\bullet$}{
In addition, SPH hardware could be developed (either as SPD or FPGA
subsystems), which would lead to an estimated cost/performance
improvement of a factor 50, for the net speed-up of simulations
including both gravity and SPH, compared to running the same
simulation with the most efficient code on the general-purpose
front-end.  }

\item{$\bullet$}{
For a full state-of-the-art simulation of the origin of large-scale
structure in the universe, additional physical processes would need to
be modeled on the front-end.  Full throughput simulations would then
achieve a net cost/performance improvement factor estimated to lie in
the range $20 \sim 30$.
}

\vfill\eject
\null\noindent
{\bf 1.  Introduction}
\medskip\noindent
This report summarizes the conclusions that were reached during a
two-day workshop on "Smooth Particle Hydrodynamics: Models,
Applications, and Enabling Technologies", at the Institute for
Advanced Study in Princeton, June 19-20, 1997.

The workshop took place in the larger context of a series of meetings,
devoted to the exploration of pathways to petaflops computing.  Of the
four main architecture categories, identified as candidates for
designing a petaflops computer during previous workshops, three have a
general-purpose character.  The fourth category pertains to
Special-Purpose Devices (SPDs).  Currently, the SPD with the highest
speed is the GRAPE-4, developed at the University of Tokyo.  The
GRAPE-4 is designed to speed up gravitational $N$-body calculations
that form the core of many simulations in astrophysics.

The name GRAPE stands for GRAvity PipE, and indicates a family of
pipeline processors that contain chips specially designed to calculate
the Newtonian gravitational force between particles.  A GRAPE
processor operates in cooperation with a general-purpose host
computer, typically a normal workstation.  The force integration and
particle pushing are all done on the host computer, and only the
inter-particle force calculations are done on the GRAPE.  Since the
latter require a computer processing power that scales with $N^2$,
while the former only require $\propto N$ computer power, load balance
can always be achieved by choosing $N$ values large enough.

The Grape-4 developers have won the Gordon Bell prize for
high-performance computing in each of the past two years.  In 1995,
the prize was awarded to Junichiro Makino and Makoto Taiji for a
sustained speed of 112 Gflops, achieved using one-sixth of the full
machine on a 128k particle simulation of the evolution of a double
black-hole system in the core of a galaxy.  The 1996 prize was awarded
to Toshiyuki Fukushige and Junichiro Makino for a 332 Gflops
simulation of the formation of a cold dark matter halo around a
galaxy, modeled using 768k particles on three-quarters of the full
machine.

In December 1995, we held a one-day workshop at NCSA, in order to
discuss possible future developments for the use of SPDs in
astrophysics.  Extending the teraflops class GRAPE-4 to a petaflops
class next-generation machine was seen to be a real possibility by the
year 2000.  In addition, we discussed ways to extend the functionality
of these SPDs beyond the simulation of purely gravitational
calculations.  In particular, the idea was suggested to make a
hardware implementation of one of the most popular algorithms for
hydrodynamics in astrophysics, SPH (Smooth Particle Hydrodynamics).

Unlike more conventional grid-based codes, an SPH code is a type of
$N$-body code, in which each particle carries with it thermodynamic
information (such as the entropy), in addition to the usual values for
its mass, position, and velocity.  Each particle determines its local
hydrodynamical conditions through an averaging procedure that involves
a number of nearest neighbors.  A variety of algorithmically different
versions of SPH exist.  For example, there is a large amount of
freedom in the choice of kernel that can be used for smoothing
purposes, in order to extract the local thermodynamic quantities from
the values carried around by neighboring particles.

At the start of the workshop, during the morning session of the first
day, the following brief review talks were presented.

\item{}{Lars Hernquist presented an overview of SPH.  He discussed the
role played by SPH modeling in large-scale simulations in
astrophysics.  He presented a comparison between different SPH
approaches, and a comparison with non-SPH versions of hydrodynamics.
}

\item{}{Thomas Sterling spoke about the future role of special-purpose
devices.  He gave a summary of the petaflops initiative and discussed
future directions for high-per\-for\-mance computing.  In addition, he
talked about the role that special-purpose devices can be expected to
play.  }

\item{}{Jeff Arnold gave an overview of the SPLASH-2 project.  This
project provided the first large-scale example of a high-speed
computer based purely on FPGAs (Field-Programmable Gate Arrays, a form
of reconfigurable logic).  He presented a brief history of the
project, discussed lessons learned from it, and mentioned some future
prospects.  }

\item{}{Piet Hut presented an overview of the GRAPE project, past and
future.  He gave a summary of the 8-year history of the project, and
listed the characteristics of the different GRAPE models.  He then
reviewed some of the science done on them, and discussed projected
future developments.  }

\item{}{Steve McMillan talked about plans and possibilities for the
GRAPE-6, the planned petaflops version of the teraflops GRAPE-4.  He
gave an outline of some ideas to combine the GRAPE-6 pure-gravity
hardware with a 1-10 Tflops FPGA system for modeling non-gravitational
physics.  }

\noindent
During the afternoon session of the first day, and throughout the
second day, various round-table discussions were held, in order to go
deeper into the questions that were raised during the presentations
listed above.  Specifically, we concentrated on the following list of
questions, provided by Thomas Sterling in his talk:

\item{$\bullet$}{
Science --- Using SPH within certain performance regimes, what new
science will be enabled.
}

\item{$\bullet$}{
Model --- To what degree is SPH a valid representation of the physical
phenomena to be modeled?  Where does it work, where will it fail?
}

\item{$\bullet$}{
Forms --- What is the range of variations of SPH that must be
incorporated and how may they be parametrized?
}

\item{$\bullet$}{
Performance --- What is the computational demand with respect to
science problem and problem size.
}

\item{$\bullet$}{
Impact --- How does computational SPH affect performance scaling of
entire science problem?
}

\item{$\bullet$}{
Data Flow --- What are the data flow rates across the SPH interface
and how does this scale with problem size and with respect to other
computational components ($N$-body simulation, etc.)
}

\item{$\bullet$}{
Implementation --- Does an SPD approach provide significant
performance advantage for SPH over alternative computational methods?
Is a reconfigurable logic methodology suitable?
}

\item{$\bullet$}{
Plan --- What is needed?  What will it cost?  How long will it take?
}

\noindent
The remainder of this report illustrate the main findings, in response
to these questions.

The scientific organizing committee for the workshop consisted of the
following six members:

\item{}{Piet Hut ({\it chair,} Institute for Advanced Study, Princeton)
}

\item{}{Lars Hernquist (Lick Observatory, University of California, Santa Cruz)
}

\item{}{George Lake (Astronomy Department, University of Washington, Seattle)
}

\item{}{Jun Makino (Department of General Systems Study, Komaba, University of
Tokyo, Japan)
}

\item{}{Steve McMillan (Department of Physics and Atmospheric Science,
Drexel University, Philadelphia)
}

\item{}{Thomas Sterling (Jet Propulsion Laboratory, Pasadena \& California
Institute of Technology, Pasadena)
}

\noindent
In total, there were 19 participants.  In additions to the ones listed
above, the following individuals attended the workshop:

\item{}{Jeff Arnold (independent consultant)
}

\item{}{Romeel Dav\'e (Astronomy Department, University of California,
Santa Cruz)}

\item{}{Kimberly Engle (Department of Physics and Atmospheric Science,
Drexel University, Philadelphia)
}

\item{}{Toshiyuki Fukushige (Department of General Systems Study, Komaba,
University of Tokyo, Japan)
}

\item{}{Neal Katz (Dept. of Physics and Astronomy, University of
Massachusetts, Amherst)
}

\item{}{Nobuyuki Masuda (Department of Earth science and Astronomy, Komaba,
University of Tokyo, Japan)
}

\item{}{Julio Navarro (Steward Observatory, University of Arizona, Tucson)
}

\item{}{Mike Norman (Department of Astronomy, University of Illinois, Urbana)
}

\item{}{Kevin Olson (George Mason Univ. and NASA/GSFC, NASA/GSFC, Greenbelt)
}

\item{}{Matthias Steinmetz (Steward Observatory, University of Arizona, Tucson)
}

\item{}{Frank Summers (Columbia Astrophysics Lab, Columbia University, New York)
}

\item{}{Peter Teuben (Astronomy Department, University of Maryland, College Park)
}

\item{}{James Wadsley (CITA/Department of Astronomy, University of Toronto,
Toronto, Canada) }

\vfill\eject
\hbox{}
\bigskip
\bigskip
\bigskip
{\bf 2.  Smooth Particle Dynamics: the Computational Challenge}
\medskip 

\par\noindent

There are a number of excellent review articles describing SPH in depth,
particularly that due to Monaghan (1992; ARA\&A 30, 543).  What follows
is a cursory survey of the method, emphasizing the challenges that face
designers of special-purpose hardware.

Smoothed particle hydrodynamics was originally formulated by Lucy
(1977; AJ 82, 1013) and Gingold \& Monaghan (1977; MNRAS 181, 375) to
model self-gravitating fluids in astrophysics.  Unlike traditional
Eulerian algorithms, SPH represents a fluid with particles and does
not require a grid to integrate the equations of motion.  This feature
makes SPH ideal for situations were a large dynamic range is needed or
systems where the ``interesting'' parts of the fluid occupy only a
small fraction of the simulation volume.  It has been applied with
success to problems ranging from astronomical impacts and stellar
collisions to galaxy mergers and the formation of large-scale
structures in the Universe.  The ease with which other physical
processes can be incorporated into SPH codes also makes this method
well-suited for modeling phenomena outside the astronomical realm,
including nuclear dynamics, MHD instabilities, the dynamics of solids,
and free surface flows.

Intuitively, SPH describes a fluid by replacing its continuum
properties with locally averaged (smoothed) quantities.  In this
sense, the formal development of SPH is not unlike the mathematical
theory of generalized functions.  Local averages are performed by
dividing the fluid into elements which carry a certain mass and
then replacing integral quantities by discrete sums over the mass
elements.  For example, the locally averaged density in SPH is
given by expressions of the form
$$
     < \rho ({\bf r} ) > \, \approx \, \sum_{j=1}^N m_j 
        W({\bf r} - {\bf r}_j , h ) \, , \eqno(1) 
$$ 
where the angle-brackets imply that the density has been smoothed over
a region whose extent is determined by the ``smoothing kernel''
$W({\bf r}, h)$ over the characteristic length-scale, $h$.  Formally,
the estimate provided by equation (1) is equivalent to summing over
$N$ particles each having a density profile determined by the
smoothing kernel.  A real fluid can also be imagined to consist of
fluid ``particles'' provided that these particles are small compared
to the scales over which macroscopic properties of the fluid varies,
but large enough to contain many molecules so that macroscopic
averages can be defined sensibly.  In an SPH computation, the number
of simulation particles is necessarily small and, hence, the
averaging scale, set by $h$, may not always be small compared to the
distance over which the fluid property varies.  This fact emphasizes
the need for large numbers of particles in the calculations, since
the exact continuum limit is recovered only as $N\rightarrow \infty$.

The procedure just described is analogous to that used to formulate a
discrete (N-body) version of collisionless dynamics.  Similar to that
development, smoothing in SPH can be extended to any physical property
and can be applied to derive equations of motion for the fluid that
reduce to the Navier-Stokes equations for a sufficiently large number
of fluid elements.  Computationally, as in an N-body algorithm, SPH
codes follow the motions of particles which, in addition to their
mass, carry the hydrodynamic and thermodynamic information needed to
specify the evolution of the fluid.  Thus, particles in SPH are like
nodes in a mesh, but one that is continuously deformable and distorts
automatically to put more of the computational effort in regions of
relatively high density.  Gradients are calculated from the smoothing
procedure, which makes it possible to interpolate between particles.

A number of practical issues must be addressed before the
functionality of SPH can be hardwired.  The choice of the smoothing
kernel is not unique, although it is desirable to employ a form such
that $W$ is sharply peaked, to preserve the local character of the
smoothing process.  In some cases, it may prove advantageous to employ
an anisotropic smoothing kernel, depending on the symmetry of the
flow.  Regrettably, there is no formal theory for error propagation in
SPH, because the rate of convergence of the smoothed estimates depend
on how the particles are distributed and how they evolve, and so it is
not possible to select the optimal smoothing kernel a priori.  (The
same would be true for a mesh-based code in which the grid points are
free to move with the fluid in an arbitrary manner.)  The discrete
form of the equations of motion is not unique and various choices lead
to forms that propagate errors slightly differently, depending on the
particle number.  Usually, it is advantageous to give each particle
its own smoothing length, and to allow smoothing lengths to vary
dynamically as the structure of the fluid changes.  In this case, the
form of the equations of motion is again not unique, and neither is
the prescription used to update smoothing lengths from one timestep to
another.  Finally, SPH requires an artificial viscosity to capture
shocks.  The best choice for the artificial viscosity is not known,
and it may, in fact, be application-dependent.

In a sense, all these ambiguities are issues of ``detail'' and it
could be the case that differences introduced by variations in things
like the smoothing kernel would be unimportant in the limit where the
particle number were multiplied by a factor of, say 100 relative to
what is currently feasible.  Perhaps a more serious problem
confronting the development of special purpose hardware for SPH is the
need to include additional physical processes and the reality that the
extra physics will vary widely from one application to another.  When
applied to cosmology and galaxy formation, for example, SPH codes must
handle the interaction between the gas and background radiation
fields, follow the ionization state of the gas, and account for the
effects of star formation and feedback.  Applications to the
interstellar medium require a treatment of magnetic fields in the gas,
account for diffusive processes on small scales, and contend with the
structure of a multi-phase medium.  Even more exotic effects may be
needed to model, for example, mergers of compact and ordinary stars,
such as nuclear burning and general relativity.  Designing hardware
that would be both computationally efficient yet sufficiently flexible
to be easily adaptable to include all these physical processes
would appear to be a formidable challenge indeed.

\vfill\eject
\hbox{}
\bigskip
\bigskip
\bigskip
{\bf 3. Issues and Opportunities}

\medskip\noindent{$\bullet$}{\it~
Science --- Using SPH within certain performance regimes, what new
science will be enabled?
}
\par\noindent
In astrophysics, most hydrodynamics applications span many orders of
magnitude in density.  This makes a straightforward application of a
non-adaptive grid code very inefficient.  3-D Lagrangian grid codes,
or codes using a hierarchy of adaptive grids, are still largely under
development.  SPH, in contrast, is fully 3-D and has been widely used
in astrophysics for more than ten years.  It is an intrinsically
adaptive algorithm, since lower densities are simply reflected in
wider particle spacing.

Current applications of SPH span a wide range of astrophysical
regimes, from planetary formation and the dynamics of interstellar
matter to the behavior of interacting galaxies and the large-scale
structure of the expanding universe.  Increasing the speed of those
calculations will benefit all these areas.  Certain critical problems
in these areas require simulations with spatial and/or temporal
resolution substantially beyond what is presently available, and even
beyond what is anticipated to become available within the next few
years.

The desired resolution in an SPH simulation determines the number of
SPH particles required.  The cost of a typical calculation (including
gravity) scales somewhat faster than the total number of particles.
Particle numbers in the range 1--$10\times 10^6$ are currently
feasible.  For problems such as encounters between individual stars or
individual galaxies, this allows fairly detailed modeling of global
behavior.  For a resolution of local detail, however, such as
interactions between individual molecular clouds in galaxy---galaxy
collisions, or galaxy structure and interactions in cosmological
simulations, more that $10^8$ particles will be needed.  Such
calculations will require computer speeds in the petaflops domain.

\medskip\noindent{$\bullet$}{\it~
Model --- To what degree is SPH a valid representation of the physical
phenomena to be modeled?  Where does it work, where will it fail?
}
\par\noindent
In the limit of very large particle number, the SPH equations
reproduce the Navier-Stokes equations of fluid dynamics.  In this
sense, SPH has been proven to be correct.  In practice, the adequacy
of SPH strongly depends on the particular problem under consideration.
For example, an accurate simulation of shocks will require a much
larger particle number than a model in which the fluid flows are more
smooth.  The method's scaling and error properties are still the
subject of active research.

A substantial increase in the number of particles in a simulation will
likely have to be accompanied by the inclusion of a significantly
greater degree of physical detail, such as radiative transfer,
magnetic fields, ionization and cooling, and a treatment of
multi-phase media.  Perhaps the most challenging problem in
large-scale simulations is to provide an adequate treatment of star
formation.  Here, the computational speed-up will help, but there
still remain several unresolved issues concerning the physical effects
involved.

\medskip\noindent{$\bullet$}{\it~
Forms --- What is the range of variations of SPH that must be
incorporated and how may they be parametrized?
}
\par\noindent
During the twenty years since the first appearance of SPH, a wide
variety of algorithms and implementations has appeared in the
literature.  It is fair to say that currently almost every individual
researcher has his or her favorite method, differing in several
aspects from those of their colleagues.  The major areas of divergence
among current SPH implementations are: (1) the choice of smoothing
kernel, (2) the choice and specification of smoothing length, (3) the
symmetrization of the equations of motion, (4) the form of the
artificial viscosity term used to handle shock propagation and (5) the
form and implementation of the energy equation.

While we discussed these differences, it became clear that
considerable speed-up would entice most SPH researchers to give up
their own favorite scheme in order to use the scheme implemented in
the fast hardware.  Therefore, the current variation in
implementations should not stand in the way of a special-purpose
implementation of SPH.  However, it was also clear that handling a lot
of ``new'' physics on the front end would substantially reduce the
overall speedup achieved.

\medskip\noindent{$\bullet$}{\it~
Performance --- What is the computational demand with respect to
science problem and problem size.
}
\par\noindent
Gravity plays an important role in all astrophysical applications of
SPH, making it impossible to separate the gravitational physics from
the hydrodynamics in any practical way.  Thus, analysis of
computational demand necessarily requires that both components of the
problem be considered together.  In many cases, the computation of the
gravitational force provides essential information (such as neighbor
lists) for the SPH calculation.  A further complication for cosmology
is the fact that periodic boundary conditions must be applied,
increasing by a non-negligible factor the cost of tree-based
gravitational force calculations.

In a ``typical'' tree-based cosmological code, the breakdown of
computational effort among the various parts of the program is as
follows: (1) tree construction and traversal (including neighbor list
determination), 20\%; (2) computation of gravitational interactions,
70\% (20\% of this associated with the implementation of periodic
boundary conditions); (3) SPH force calculation, 10\%.

We will analyze the impact of SPDs on these components separately,
although we recognize that separating the functionality into two or
more distinct hardware components may incur a substantial bandwidth
penalty in communicating results among the various parts of the
system.  Note that it makes sense to speed up the SPH calculation only
if steps are first taken to accelerate the gravity and tree-handling
portions of the code.

\medskip\noindent{$\bullet$}{\it~
Impact --- How does computational SPH affect performance scaling of
entire science problem?
}
\par\noindent
In currently existing codes, running on workstations or moderately
parallel machines, SPH implementations are highly intertwined with the
treewalk part of the gravity calculations.  Together, they make up
roughly one third of the computational expense.  If the gravity part
of the calculation were sped up by the use of SPDs, the SPH
calculations would become the bottleneck, limiting the overall
speed-up to only one order of magnitude, irrespective of how fast the
gravitational calculations could be performed.

If, however, the SPH part could be implemented in hardware, in SPD or
FPGA form, then the net speed-up would be significantly greater.  The
overall speed would then be determined by the next bottleneck---the
additional physics, such as radiative transport and cooling, mentioned
above.  More detailed estimates are presented in the table below.

\medskip\noindent{$\bullet$}{\it~
Data Flow --- What are the data flow rates across the SPH interface
and how does this scale with problem size and with respect to other
computational components ($N$-body simulation, etc.)
}
\par\noindent
Considering just the SPH subsystem, calculation of the SPH portion of
the force on a single particle requires that $\sim$100 neighbor
interactions be computed, at a cost of about 100 floating-point
operations per interaction.  The total amount of particle data
required (for the particle under consideration, not its neighbors) is
on the order of 100 bytes.  The minimum bandwidth thus is $\sim$0.01
bytes/flop, or 100 Gbyte/s for a 10 Tflops SPH system.  However, this
calculation assumes that there is no overhead involved in obtaining
and maintaining the required neighbor information, which is
unrealistic.  There was no discussion of ways in which neighbor
information might be transferred efficiently between the gravity and
the SPH subsystems.

\medskip\noindent{$\bullet$}{\it~
Implementation --- Does an SPD approach provide significant
performance advantage for SPH over alternative computational methods?
Is a reconfigurable logic methodology suitable?}
\par\noindent
For the pure SPH problem, it appears that a hardware implementation of
SPH could produce a substantial improvement in speed, although the I/O
issues just described remain unresolved.  Whether an algorithm as
complex as a typical SPH force loop can be implemented in hardware is
unclear.  Whether custom or reconfigurable hardware is chosen, it
appears that a significant reduction in precision would be necessary
in order to fit an SPH pipeline into a chip.  The possibility of using
multiple FPGAs to implement a single pipeline was discussed briefly,
and may be feasible.

There was disagreement as to whether reduced (i.e.~less than 32-bit)
precision was adequate.  The statistical errors inherent in using only
100 neighbors to determine fluid quantities seem to imply that only a
few decimal digits of accuracy are actually needed; however, some SPH
users maintained that (presently unimplemented) SPH applications
involving magnetic fields would require full double precision.  No
consensus was reached on this issue.

\medskip\noindent{$\bullet$}{\it~
Plan --- What is needed?  What will it cost?  How long will it take?
}
\par\noindent
For a 100 Tflops pure gravity engine, the estimated cost is $\sim$\$2
million.  The cost of a 10 Tflops SPH subsystem implemented in
reconfigurable hardware would probably be similar.  The gravity engine
has been funded by the Japanese government and should be completed by
2000.  Implementation of SPH in FPGAs could possibly be carried out in
a similar timeframe, but only with significant input from the SPH user
community.

An important side effect of the construction of a 100/10 Tflops
gravity/SPH system would be the availability of single-board
workstation accelerators for use by individual researchers, offering
1/0.1 Tflops performance at modest cost (less than \$50,000, say).
However, the development costs of the gravity and SPH hardware cannot
be justified on the basis of these small systems alone.

\bigskip\bigskip\bigskip\bigskip\noindent
{\bf 4. Findings}
\medskip\noindent
As mentioned above, the total computational cost of a typical
cosmological SPH code using a Barnes-Hut tree for the gravity
computation, consists of several basic components: computation of
pairwise gravitational interactions (with cost scaling as $O(N)$,
where $N$ is the number of particles involved), tree construction and
traversal (reducing the gravitational cost to $O(\log(N))$), inclusion
of periodic boundary conditions (the so-called ``Ewald'' method), and
the computation of fluid properties using SPH.  The workshop's
conclusions on the likely speed-up that might be achieved by
hardwiring some or all of these components are summarized in the table
below.

$$\vbox{\tabskip=0pt \offinterlineskip
\halign to 365pt{\strut#& \vrule#\tabskip=2em plus2em& \hfil#&
\vrule#& \hfil#\hfil&
\vrule#& \hfil#\hfil&
\vrule#& \hfil#\hfil&
\vrule#& \hfil#\hfil&
\vrule#& \hfil#&
\vrule#\tabskip=0pt\cr \noalign{\hrule}
& & \omit\hidewidth \  \hidewidth & & \omit\hidewidth
$O(N)$ \hidewidth & & \omit\hidewidth Tree 
\hidewidth & & \omit\hidewidth Ewald \hidewidth & & \omit\hidewidth SPH
\hidewidth & & \omit\hidewidth Full \hidewidth & \cr \noalign{\hrule}
& & $O(N)$ & & 1,000 & & 20 & & 5 & & 2 & & 2 & \cr \noalign{\hrule}
& & Tree & & \  & & 100 & & 5 & & 2 & & 2 & \cr \noalign{\hrule}
& & Ewald & & \  & & \  & & 100 & & 5 & & 4 & \cr \noalign{\hrule}
& & SPH & & \  & & \  & & \  & & 50 & & 20 & \cr \noalign{\hrule}
}}$$

The top line of the table lists the various applications.  (The final
application, simply labeled ``Full,'' assumes that more complex
physics, such as heating and cooling, and radiative transport, are
also included.)  The list is cumulative---each item assumes the use of
all items to its left.  The left-most column lists the part of the
calculation that is presumed to be hard-wired.  Again, this list is
cumulative---each row presumes that all rows above it are hard-wired.
The numbers indicate the improvement in performance that might be
achieved over a state-of-the-art general-purpose system in 2001, using
the custom and reconfigurable devices expected to be available in that
year, operating in tandem with a general-purpose host.

\vfill\eject
\hbox{}
\bigskip
\bigskip
\bigskip
{\bf 5.  Conclusions and Recommendations}
\medskip\noindent
\item{1.}{Hardwiring SPH is feasible, if we can accept a standard
algorithm and can live with reduced precision.  The former appears to
be the case, the latter is less clear.  We note also that SPH is
probably the most complex astrophysical application we would consider
implementing with the technology likely to be available within the
next five years.}

\item{2.}{The speedup achieved would be significant, but not
overwhelming.  Furthermore, it can be achieved only if other more
costly components of the calculation are handled first.}

\item{3.}{Significant challenges exist in designing the
architecture of a petaflops-class system combining special-purpose
gravitational and SPH components.  In particular, the dataflow problem
associated with neighbor interactions is serious and unresolved.}

\item{4.}{While considerable interest was expressed by SPH users in
the possibility of obtaining teraflops-class systems derived from the
larger system, the significant development time and expense of both
hardware and software mean that such applications alone do not provide
sufficient grounds for this project.}

\item{5.}{It is futile to undertake such an extensive  project without
the active participation of at least some part of the potential user
base.  At present, there does not seem to be a critical mass within
the SPH community to go this route.}

\item{6.}{Hardwiring the $\log N$ gravity part will give very
significant improvement in cost-performance ratio, of order a factor
100 over general-purpose high-per\-for\-mance computers.  This may be
a better route to take, leaving SPH to be computed on the front end,
and opening up the possibility of extremely fast execution of a large
number of non-SPH pure-gravity problems.}

\bigskip\bigskip\bigskip
\noindent
Acknowledgments: This workshop was sponsored by NASA, through
Dr. James Fischer, Project Manager, NASA HPCC/ESS Project, and the
Universities Space Research Association, at Goddard Space Flight
Center.

\bye